# Frequency splitting in undulator radiation from solid-state crystalline undulator

A. Shchagin,[a, b, 1] G. Kube,[a] A. Potylitsyn,[c] and S. Strokov[a]

[a] *Deutsches Elektronen-Synchrotron DESY,*
  *Notkestr. 85, 22607 Hamburg, Germany,*
[b] *Kharkiv Institute of Physics and Technology,*
  *Academicheskaya 1, Kharkiv 61108, Ukraine,*
[c] *Institute of Applied Problems of Physics,*
  *25, Hr. Nersisyan Str., 0014, Yerevan, Republic of Armenia*
  *[1]E-mail: alexander.shchagin@desy.de*

ABSTRACT: The set of frequencies and angular properties of radiation emitted from a solid-state crystalline undulator based on the channeling effect are considered. High-frequency and low-frequency branches of the undulator radiation and the angular distribution of the emitted radiation are analyzed. The ranges of frequencies and angles of radiation emitted from the solid-state crystalline undulator based on the channeling effect are found. The frequency splitting and the energy threshold in the production of undulator radiation are shown and discussed.

---

[1] Corresponding author.

**Contents**

**1. Introduction**

Classic magnetic undulators can generate X-ray beams with photon energies up to hundreds of keV, as seen in, e.g., in ref. [1]. To generate radiation in the MeV range, one has to use an undulator with a shorter period. A much shorter period can be provided by the application of a crystalline undulator.

The solid-state crystalline undulator, based on the effect of charged particles channeling in a periodically bent crystal, was proposed in refs. [2, 3]. Experimental research on the solid-state crystalline undulator began with a 10 GeV positron beam [4] and on electron beams with energies of 270 and 855 MeV in refs. [5, 6]. The crystalline undulator with straight sections based on the same channeling effect was proposed in ref. [7]. The crystalline undulator based on the volume reflection effect was briefly proposed at the end of ref. [8] and some of its properties are recently considered in [9]. References for numerous theoretical studies on the properties of crystalline undulators can be found, e.g., in papers [7, 10-14, 19].

Here, we consider the spectral and angular properties of undulator radiation emitted from a solid-state crystalline undulator based on the channeling effect, taking into account the radiation emission in the crystalline medium. The splitting of the radiation frequency and the energy threshold for undulator radiation emitted from a crystalline undulator are shown and discussed.

**2. Calculations**

Properties of radiation emitted by relativistic particles in a medium differ from those emitted in vacuum. The equation for the angular frequency ω of radiation emitted in a periodical crystalline medium was derived by Ter-Mikaelian in his monograph [15]:

$$\omega = \frac{2\pi n}{l} \frac{V}{1 - \frac{\sqrt{\varepsilon}}{c} V \cos\theta}, \quad (1)$$

where $V$ is the velocity of the particle moving rectilinearly in a periodical structure with period $l$, $\varepsilon = 1 - \left(\frac{\omega_p}{\omega}\right)^2$ is the permittivity of the medium (considered for X- and gamma-ray ranges), $\omega_p = \sqrt{\frac{e^2 n_e}{m \varepsilon_0}}$ is the plasma frequency in the medium, $e, m, n_e$ are electron charge, mass, and density respectively, $\varepsilon_0$ is the vacuum permittivity, $n$ is the harmonic number, $c$ is the speed of light, and $\theta$ is the polar observation angle with respect to the incident particle velocity vector $\vec{V}$. Equation (1) was derived by Ter-Mikaelian [15] for parametric X-ray radiation of relativistic particles moving in a crystal based on the energy and momentum conservation laws in a crystal. The derivation of Eq (1) from Huygens' construction can be found in refs. [9, 16]. Eq. (1) was confirmed experimentally for parametric X-ray radiation emitted at an observation angle $\theta \gg \gamma^{-1}$ for example in refs. [16, 17]. Here $\gamma$ is the relativistic Lorentz factor of the incident particle, i.e. $\gamma = \left[1 - \left(\frac{V}{c}\right)^2\right]^{-\frac{1}{2}}$. In the following the focus is on high energetic radiation emitted in forward direction at $\theta < \gamma^{-1}$ and $\theta \sim \gamma^{-1}$ by a relativistic charged particle in a crystalline undulator under the condition that the energy of the emitted quanta $\hbar\omega$ is much smaller than the energy of the incident particle $E$:

$$\hbar\omega \ll E. \tag{2}$$

The average longitudinal velocity $V_z$ of the particle in sinusoidal motion in a vacuum or crystalline undulator is smaller than the particle velocity $V$. Therefore, one has to use the longitudinal velocity in Eq. (1). The average longitudinal velocity in sinusoidal motion is expressed using the undulator parameter $K$ (see, e.g., Eq. (5.25) in [1])

$$V_z = c\left(1 - \frac{1 + \frac{K^2}{2}}{2\gamma^2}\right), \tag{3}$$

with $K$ as the ratio of the maximum angle $\alpha$ between particle velocity vector and undulator axis to $\gamma^{-1}$, i.e. $K = \alpha\gamma$ [1]. For sinusoidal motion of the particle with period $l$ and amplitude $A$, this angle is $\alpha = \frac{2\pi A}{l}$. Therefore,

$$K = \frac{2\pi A}{l}\gamma. \tag{4}$$

The transverse motion of the particle due to channeling oscillations (see, e.g., Eqs. (17-19) in ref. [13]) can lead to some increase of the undulator parameter [13]. The device works similar to an undulator and emits coherent radiation for $K < 1$, and similar to a wiggler for $K > 1$ [1]. In the following we will consider mainly the case of coherent undulator radiation at $K < 1$.

For the derivation of the radiation frequency based on Eq. (1) one has to use the expression for the average longitudinal velocity (3) instead of $V$. Assuming ultra-relativistic particles in the following, the relations $\alpha$, $\theta$, $\frac{\omega_p}{\omega}$, $\gamma^{-1} \ll 1$ holds for the corresponding parameters. Neglecting higher order terms, one obtains from Eq. (1) the undulator equation for the frequency of radiation emitted from a crystalline undulator based on the channeling effect

$$\omega = \frac{4\pi n c}{l} \gamma^2 \frac{1}{\left[1 + \frac{K^2}{2} + (\gamma\theta)^2 + \left(\frac{\gamma\omega_p}{\omega}\right)^2\right]}. \tag{5}$$

The equation (5) differs from the well-known undulator equation in vacuum [1] by the term $\left(\frac{\gamma\omega_p}{\omega}\right)^2$ which appeared because of the increase in the phase light velocity $\frac{c}{\sqrt{\varepsilon}}$ in the crystal compared to the light velocity $c$. More detailed discussion about this subject can be found in ref. [9].

The undulator equation (5) is a quadratic one which can be rewritten in the form

$$\omega^2\left(1 + \frac{K^2}{2} + (\theta\gamma)^2\right) - \omega \frac{4\pi n c}{l} \gamma^2 + (\gamma\omega_p)^2 = 0. \tag{6}$$

The solutions of equation (6) are the radiation frequencies $\omega$ from a solid-state crystalline undulator based on the channeling effect

$$\omega = \frac{\omega_0}{2} \pm \sqrt{\left(\frac{\omega_0}{2}\right)^2 - \frac{(\gamma\omega_p)^2}{1 + \frac{K^2}{2} + (\gamma\theta)^2}}, \tag{7}$$

where

$$\omega_0 = \frac{4\pi n c}{l} \gamma^2 \frac{1}{\left[1 + \frac{K^2}{2} + (\gamma\theta)^2\right]} \tag{8}$$

is the "vacuum" solution of Eq. (5) at $\omega_p = 0$. The "vacuum" solution (8) is valid for a magnetic undulator where the particles emit radiation in vacuum [1]. However, Eq. (8) is sometimes used in estimations of the radiation frequency from crystalline undulators [5, 13, 14, 18].

    The properties of radiation emitted in a solid-state crystalline undulators were studied in refs. [10, 11]. Expressions similar to Eq. (7) were stated in both works, however being independent on the observation angle $\theta$ (see Eq. (13) in Ref. [10] and Eqs. (28, 29) in Ref. [11]).

    The frequency $\omega$ in Eq. (7) as a function of the relativistic factor $\gamma$ of the incident particle can have two, one or no solution depending on sign and value of the expression under the root. Undulator radiation exists under the condition that the radicand in Eq. (7) is non-negative:

$$\left(\frac{\omega_0}{2}\right)^2 \geq \frac{(\gamma\omega_p)^2}{1+\frac{K^2}{2}+(\gamma\theta)^2}. \tag{9}$$

From the inequality (9) and the expression for the undulator parameter (4), one can find that the radiation is possible at the condition if the relativistic factor $\gamma$ exceeds or is equal the threshold value $\gamma_{thr}$

$$\gamma \geq \gamma_{thr}, \tag{10}$$

where

$$\gamma_{thr} = \frac{1}{\sqrt{\left(\frac{2\pi nc}{l\omega_p}\right)^2 - 2\left(\frac{\pi A}{l}\right)^2 - \theta^2}}. \tag{11}$$

The minimum threshold $\gamma_{thr}^{\min}$ occurs at zero observation angle $\theta = 0$

$$\gamma_{thr}^{\min} = \frac{1}{\sqrt{\left(\frac{2\pi nc}{l\omega_p}\right)^2 - 2\left(\frac{\pi A}{l}\right)^2}}. \tag{12}$$

The same threshold (12) at zero observation angle was found in refs. [10, 11], see Eq. (16) in ref. [10] and Eq. (30) in ref. [11]. But the threshold $\gamma_{thr}$ increases at increasing observation angle $\theta$:

$$\gamma_{thr} = \gamma_{thr}^{\min}\sqrt{(\theta\gamma)^2 + 1}, \tag{13}$$

where term $\theta\gamma$ is the observation angle $\theta$ in $\gamma^{-1}$ units.

The range of observation angles $\theta$ where radiation exists is restricted:

$$0 \leq |\theta| \leq \theta_{\max}, \tag{14}$$

where the value $\theta_{\max}$ can be found from inequality (9)

$$\theta_{\max} = \sqrt{\left(\frac{2\pi nc}{l\omega_p}\right)^2 - \frac{1+2\left(\frac{\pi A}{l}\right)^2\gamma^2}{\gamma^2}}. \tag{15}$$

According to Eq. (15), the angle $\theta_{\max}$ as a function of $\gamma$ increases from zero at $\gamma = \gamma_{thr}^{\min}$ to the value $\theta_{\max 1}$ at $\gamma \to \infty$:

$$\theta_{\max 1} = \sqrt{\left(\frac{2\pi nc}{l\omega_p}\right)^2 - 2\left(\frac{\pi A}{l}\right)^2}. \tag{16}$$

One can write the relation between $\gamma_{thr}^{\min}$ (12) and $\theta_{\max 1}$ (16):

$$\theta_{\max 1}\gamma_{thr}^{\min} = 1. \tag{17}$$

The solution (7) of the undulator equation (5) provides two branches of radiation frequencies from the channeling-based solid-state crystalline undulator at $\gamma > \gamma_{thr}$: the low-frequency solution (or branch) $\omega_1$ for the negative sign in Eq. (7) and the high-frequency solution (or branch) $\omega_2$ for the positive sign in Eq. (7). Both branches have asymptotes at $\gamma \gg \gamma_{thr}$. They can be found from Eq. (7) under the condition $\left(\dfrac{\omega_0}{2}\right)^2 \gg \dfrac{(\gamma\omega_p)^2}{1+\dfrac{K^2}{2}+(\gamma\theta)^2}$. The low-frequency asymptote $\omega_1^{as}$ is

$$\omega_1^{as} = \dfrac{l\omega_p^2}{4\pi nc}. \qquad (18)$$

It is independent of incident particle energy and observation angle. The harmonic frequencies in Eq. (18) are inversely proportional to the harmonics number $n$. The asymptote of the high-frequency solution $\omega_2^{as}$ is

$$\omega_2^{as} = \omega_0 - \omega_1^{as} = \dfrac{4\pi nc}{l}\gamma^2 \dfrac{1}{\left[1+\dfrac{K^2}{2}+(\gamma\theta)^2\right]} - \dfrac{l\omega_p^2}{4\pi nc}, \qquad (19)$$

where the undulator parameter is determined by Eq. (4). The high-frequency asymptote (19) approaches the "vacuum" solution (8) as the harmonic number $n$ increases.

For illustration of the calculations, we used parameters of the Si crystalline undulator (sample #1) which has been produced and described in ref. [18]. The undulator is designed for experimental research of the undulator radiation on 10 GeV positron beam [18]. The undulator has period $l = 334$ micrometers and amplitude $A = 1.28$ nanometres. The value $\hbar\omega_p = 31.1$ in silicon single-crystal. Eq. (7) is a two-valued function of two arguments, where it exists. The results of calculations of emitted photons energy $\hbar\omega$ as a function of positron or electron energy $E = mc^2\gamma$ and $\theta\gamma$, which is the observation angle $\theta$ in $\gamma^{-1}$ units, are shown in Fig. 1.

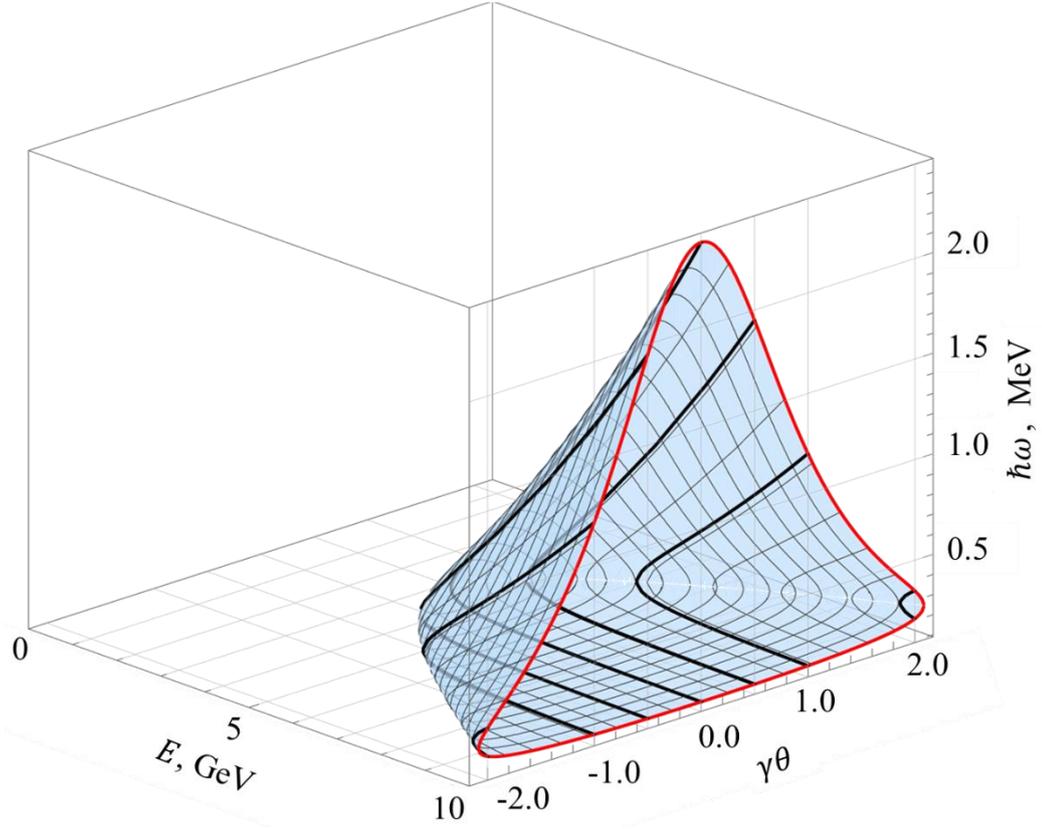

**Figure 1.** The undulator radiation photon energy $\hbar\omega$ as a function of the positron or electron energy $E$ and emission angle $\theta$ in $\gamma^{-1}$ units. Calculations are performed by Eq. (7) for Si crystalline undulator "sample #1" described in ref. [18] for harmonic number $n=1$. Photon energies of radiation emitted at $\theta\gamma = 0.5, 1.0, 1.5, 2.0$ are shown by black thick curves. Photon energies emitted at positron or electron energy $E=10$ GeV are shown by red curve.

The threshold nature of undulator radiation from the crystalline undulator is well seen in Fig. 1. The minimum threshold energy is $mc^2\gamma_{thr}^{min} = 4.325$ MeV at observation angle $\theta\gamma = 0$ (see Eq. (12)). At particles energy E<4.325 MeV the undulator radiation is impossible. The threshold energy increases at increasing the observation angle, see Eqs. (11,13) in Fig. 1. The calculations shown in Fig. 1 were performed for point-like detector. But in real experiments, any detector has restricted angular size that usually is determined by a collimator angular size. The collimator angular size usually is expressed in $\gamma^{-1}$ units. Results of calculations of the photon energies that can be registered by a detector collimated by round collimator with angular size (angular radius) $\theta\gamma = 0.5$ installed in the undulator axis are shown in Fig. 2.

Calculations of the photon energies of undulator radiation $\hbar\omega$, the "vacuum" solution $\hbar\omega_0$, the low-frequency asymptote $\hbar\omega_1^{as}$, and the high-frequency asymptote $\hbar\omega_2^{as}$ as functions of the positron or electron energy $E = mc^2\gamma$ for a solid-state Si undulator with a period of 334 µm and amplitude $A = 1.28$ nm (sample #1 in ref. [18]) are shown in Fig. 2. Calculations were performed for $\theta\gamma$ equal to 0 and 0.5. Also, the results of the calculation of the cutoff energy

$$\hbar\omega_{cf} = \gamma\hbar\omega_p \qquad (20)$$

are presented in Fig. 2. The role of the cutoff energy is discussed in the "Discussion" section.

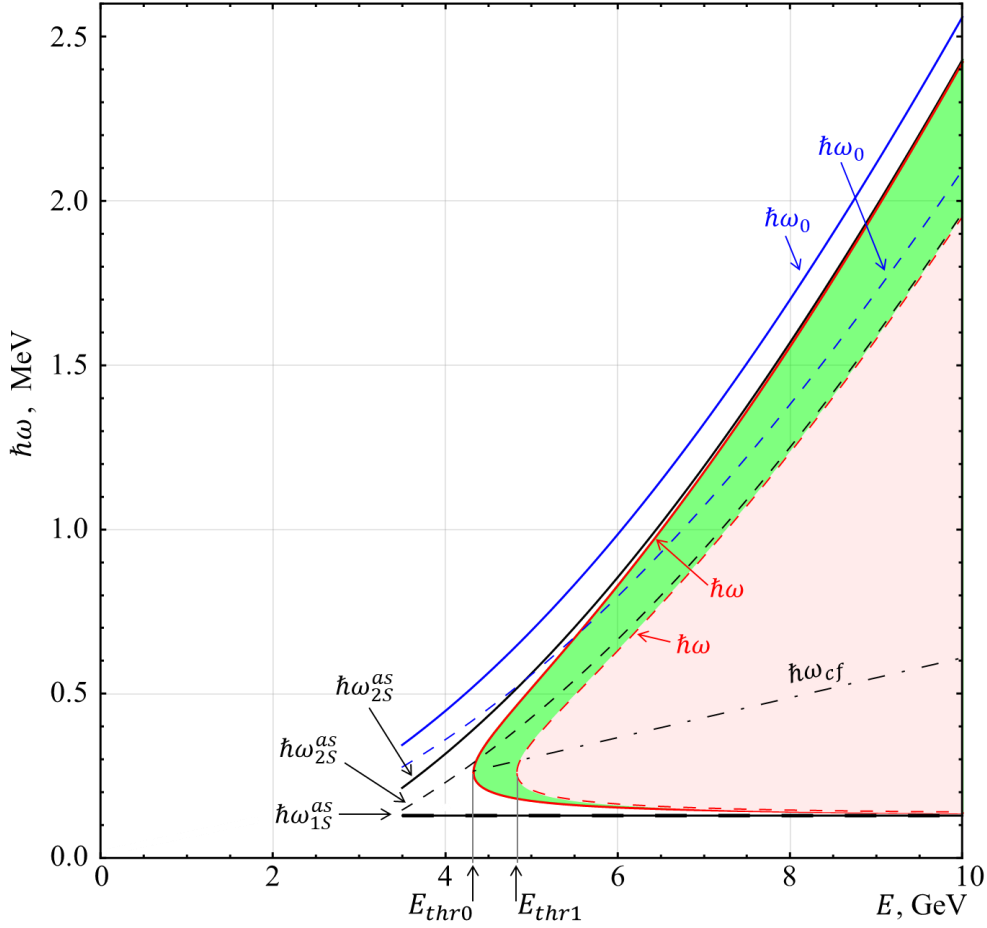

**Figure 2.** Photon energies for a solid-state silicon undulator based on the channeling effect are plotted as functions of the positron or electron energy $E$. The silicon undulator has a period of $l = 334$ µm and an amplitude of $A = 1.28$ nm. Parameters include the harmonic number $n = 1$ and $\hbar\omega_p = 31.1$ eV. Calculations of the values $\hbar\omega$, $\hbar\omega_0$, $\hbar\omega_1^{as}$, $\hbar\omega_2^{as}$ are performed using Eqs. (7, 8, 18, 19) respectively. Calculations at observation angle $\theta = 0$ are depicted by solid lines. The same calculations at an observation angle of $\theta = 0.5\gamma^{-1}$ are represented by dashed lines. Both calculations of $\hbar\omega$ are shown by red curves. The area with photon energies which can be observed by a collimated detector with angular size $0.5\gamma^{-1}$ is shown in green color. This area is between two red curves $\hbar\omega$. The areas prohibited for undulator radiation observation are shown

in pink color inside the $\hbar\omega$ curves and in white color outside the $\hbar\omega$ curves. Additionally, the cutoff energy (20) is shown by a dash-dotted line. The thresholds calculated by Eqs. (11-13) are shown by two arrows $E_{thr0} = 4.325$ GeV and $E_{thr1} = 4.835$ GeV for observation angles $\theta = 0$ and $\theta = 0.5\gamma^{-1}$ respectively. One can see from Fig. 2 that low-frequency (18) and high-frequency (19) asymptotes can be used for estimations of photon energies at electron or positron energy some exceeding the threshold value.

The observation of undulator radiation from the crystalline undulator is possible, if photon energy is in green area show in Fig. 2. For example, at particle energy 10 GeV, the radiation in the high-energy branch is possible in the energy range 1.950 - 2.422 MeV with full width 0.472 MeV and relative width about 0.22. This range is in the right edge of the wide spectral peak in Fig. 4 from ref. [18], where authors of ref. [18] demonstrated results of their simulated radiation emission probability for 10 GeV positrons interacting with sample #1. The radiation in the low-energy branch is possible in the energy range 137.7 – 139.6 keV with full width 1.9 keV and relative width about 0.014. This energy range is in the left edge of the wide spectral peak shown in Fig. 4 of ref. [18]. Thus, two spectral peaks of undulator radiation with average energies 2.186 MeV in the high-frequency branch and 138.6 keV in the low-frequency branch should be observed in the experiment due to the splitting.

## 3. Discussion

Note the splitting of the frequency $\omega_0$ (8), that can be emitted in a vacuum undulator, into two frequencies $\omega_1$ (low-frequency branch) and $\omega_2$ (high-frequency branch) (7) in a crystalline undulator. The splitting indicates that two photons with energies $\hbar\omega_1$ and $\hbar\omega_2$ can be emitted from a crystalline undulator instead of only one photon with energy $\hbar\omega_0$ from a vacuum undulator with the same geometrical parameters. The energies of photons for the fixed harmonics number $n$ are (see Eq. (7))

$$\hbar\omega_1 + \hbar\omega_2 = \hbar\omega_0 \qquad (21)$$

due to the splitting. Two spectral peaks of the undulator radiation should be observed in the spectrum of the undulator radiation from a crystalline undulator.

Besides undulator radiation, other well-known mechanisms of radiation by a relativistic particle can occur in a crystalline undulator, including common bremsstrahlung, coherent bremsstrahlung on the crystalline structure, transition radiation at the surfaces of the undulator, transition X-ray radiation diffracted at the entrance crystal surface, channeling radiation, and parametric X-ray radiation from a crystalline structure. The parametric X-ray radiation is emitted at a large angle with respect to the particle velocity vector [20] and can be utilized for independent control of the undulator in the particle beam. The intensity of the radiations emitted inside the crystal, such as undulator radiation, common and coherent bremsstrahlung, channeling radiation, and parametric X-ray radiation, should be suppressed due to the Ter-Mikaelian density effect at photon energies below the cutoff energy, at $\hbar\omega < \hbar\omega_{cf}$ [15]. Conversely, the intensity of transition radiation and diffracted X-ray transition radiation should be suppressed at photon energies above the cutoff energy, at $\hbar\omega > \hbar\omega_{cf}$ [15]. The calculation of the cutoff energy is shown in Fig. 2, which reaches 609 keV at a beam particles energy of 10 GeV. The intensity of radiation in the low-frequency branch of the undulator radiation should also be suppressed due to the Ter-Mikaelian density effect because the photon energies in the low-frequency branch are below the cutoff energy (see Fig. 2 for sample #1 from ref. [18]).

One of the problems in the realization of the crystalline undulator based on the channeling effect is a short dechanneling length in a crystalline undulator. The dechanneling length of electrons is much less in comparison to one of positrons as it was discussed in, e.g., ref. [21]. Positron beam can be used for production of undulator radiation in crystalline undulator. For example, the sample #1 described in ref. [18], can operate in the undulator regime at positron beam energy up to 21.2 GeV (when the undulator parameter (4) becomes equal to unity, K=1) and in a wiggler regime at positron beam energy above 21.2 GeV.

The physical reason for both effects – the appearance of the energy threshold and the undulator frequency splitting – is the presence of a medium characterized by the plasma frequency $\omega_p$ and related exceeding of the phase velocity of light $\frac{c}{\sqrt{\varepsilon}}$ in the medium over the common light velocity $c$. Both effect disappear in vacuum, when $\omega_p = 0$: the phase velocity of light becomes equal to $c$, the energy threshold (11) becomes equal to zero, and only one undulator frequency $\omega_0$ (7,8) can be emitted.

## Acknowledgments

This project has received funding through the MSCA4Ukraine project ION-LOSS #1233244, which is funded by the European Union.